\documentclass[preprint,aps]{revtex4}
\usepackage{hyperref}
\usepackage{graphicx}
\usepackage{dcolumn}
\usepackage{bm}
\usepackage{mhchem}
\usepackage{multirow}
\usepackage{amssymb}
\usepackage{amsmath}
\usepackage{longtable}

\begin{document}


\title{Altermagnetism and Anomalous Transport in Ag$^{2+}$ Fluorides: KAgF$_3$ and K$_2$AgF$_4$}

%
%
%
\author{
  Xiao Nan Chen, Sining Zhang, Zhengxuan Wang, Minping Zhang and Guangtao Wang\footnotemark[1]
}
\date{}
\renewcommand{\thefootnote}{\fnsymbol{footnote}} 

\affiliation{%
  College of Physics, Henan Normal University, \\
  Xinxiang, Henan 453007, People's Republic of China%
}

\date{\today}
             
\begin{abstract}
Compounds containing Ag$^{2+}$ ion with 4d$^9$ configuration will cause significant  Jahn-Teller distortions and orbital ordering. Such orbital order is closely related to the magnetic coupling, according to Goodenough-Kanamori Ruels. Our first-principles calculations reveal that the ground state of KAgF$_3$ exhibits collinear A-type antiferromagnetic (A-AFM) ordering accompanied by C-type orbital ordering. In contrast, K$_2$AgF$_4$ adopts a collinear intralayer antiferromagnetic configuration coupled with ferromagnetic orbital ordering. The A-AFM  KAgF$_3$ presents distinct altermagnetic responses, including: (i) prominent anomalous transport effects, such as anomalous Hall conductivity (AHC), anomalous Nernst conductivity (ANC), and thermal anomalous Hall conductivity (TAHC); and (ii) strong magneto-optical responses, manifested through pronounced Kerr and Faraday effects. On the other hand, K$_2$AgF$_4$ behaves as a conventional collinear antiferromagnet preserving $\mathcal{PT}$ symmetry, hence precluding the emergence of an anomalous Hall response.
\end{abstract}

\maketitle

\footnotetext[1]{\href{wangtao@htu.cn}{wangtao@htu.cn}}


\section{Introduction}
Altermagnetism, a recently identified class of collinear magnetism, combines the essential characteristics of both antiferromagnets and ferromagnets, and has attracted considerable attention in condensed matter physics~\cite{Altermagnetic1,Altermagnetic2,01,02,03,04,05,06,yao3,PRB-Turek,model,model2}. Despite possessing zero net magnetization, altermagnets exhibit pronounced anomalous transport phenomena~\cite{ruo2ahc1,ruo2ahc2,ANE,ATHE,AHE1,AHE2,AHE3,AHE4,AHE5,AHE6,AHE7,AHE8,yaoAHC,yaoMOKE}, including the anomalous Hall conductivity~\cite{ANE,ATHE,AHE1,AHE2,AHE3,AHE4,AHE5,AHE6,AHE7,AHE8,yaoAHC,yaoMOKE,Ebert,Yao2025}, anomalous Nernst conductivity~\cite{ANE,yaoAHC}, and  thermal anomalous  Hall conductivity~\cite{ATHE,yaoAHC}, along with magneto-optical responses including the Kerr and Faraday effects~\cite{yaoMOKE}.Traditionally, these effects were thought to require ferromagnetism, incompatible with antiferromagnets’ zero net magnetization. However, altermagnetic materials possess a unique symmetry where the connecting operation between the two magnetic sublattices is not the conventional space-inversion ($\hat{P}$) combined with time-reversal ($\hat{T}$), but rather a rotational or mirror symmetry operation coupled with time-reversal (e.g., $R\hat{T}$ or $M\hat{T}$)~\cite{Altermagnetic1,Altermagnetic2,yao3,TSC1}. This special symmetry results in  anisotropic charge distributions in  the real-space and  band splitting in the momentum-space, distinguishing altermagnets from conventional antiferromagnets~\cite{Weyl,Topandoptical,MnTe-ARPES,shc1}.

In altermagnets, the magnetic ions with opposite spin orientations  exhibit distinct ligand-field environments from surrounding anions, which are symmetry-related through rotational or mirror operations~\cite{Altermagnetic1,Altermagnetic2}. For transition-metal oxides/fluorides with partially filled $d$-electron states, this configuration naturally combines Jahn-Teller (JT) distortions, orbital and magnetic ordering~\cite{KAgF1,KAgF2,KAgF3,YVO,YVO-OPT}. All the above properties  create the fundamental characters of altermagnetism: spin-dependent band splitting with $k$-dependent spin polarization~\cite{alter-orbital-order,alt-layer,hjzhang,shc1,giant-splitting}. So, it is proposed that many transition metal compounds would exhibit altermagnetic properties~\cite{01,02,03,04,05,06,Krempasky2024-Nature,Mazin-2023}.

The most definitive approach to verify altermagnetism consists of two essential experimental demonstrations: (i) neutron diffraction confirming the antiferromagnetic order, and (ii) spin-resolved angle-resolved photoemission spectroscopy (SR-ARPES) revealing the $k$-dependent spin splitting in the  band structure~\cite{MnTe-ARPES,RuO2-nosplitt}. However, these measurements face significant experimental challenges, particularly the demanding requirements for both high energy and spatial resolution. These challenges may lead to divergent or even conflicting conclusions among different research groups studying the same material. For instance, some studies~\cite{yaoMOKE,yaoAHC,spin-charge} have reported alter-magnetic spin moments and spin-splitting band structure in RuO$_2$, whereas others~\cite{nomagnetic,absenceAFM} have observed neither signatures of long-range magnetic order nor  spin-splitting band. To circumvent the challenge of directly proving altermagnetism, researchers often resort to indirect experimental signatures. These include measurements of the anomalous Hall effect~\cite{AHE2,AHE3,AHE4,AHE5,AHE6,AHE7,AHE8,yaoAHC,ruo2ahc1,ruo2ahc2,RuO2-nosplitt}, the magneto-optical Kerr effect~\cite{yaoMOKE}, and Faraday rotation~\cite{yaoMOKE,Chen_2024-Mn-rev}, which can provide compelling evidence for the unconventional electronic and magnetic structure characteristic of altermagnets.

The compounds containing \ce{Ag^{2+}} ions with a \(4d^{9}\) electronic configuration induce significant Jahn-Teller distortions and orbital ordering, owing to the half-filled \(e_g\) orbitals. The insulating and antiferromagnetic properties of \ce{KAgF3} and \ce{K2AgF4} have been previously reported~\cite{AgF2009,KAgF2013,KAgF1,KAgF3}. However, a comprehensive discussion is required to elucidate the complex interplay among electron correlations, orbital ordering, and super-exchange interactions that underpin this behavior.

In our study, we computed the electronic structures and total energies of \ce{KAgF3} and \ce{K2AgF4} across various magnetic configurations. The orbital ordering of \ce{Ag^{2+}} was confirmed via partial density of states (PDOS) and charge density analyses. Without incorporating electron correlation effects on the Ag \(4d\) states, both compounds exhibited negligible magnetic moments on the \ce{Ag^{2+}} ions. Upon including electron correlation parameters (\(U\)), the magnetic moments of \ce{Ag^{2+}} increased and reached values consistent with experimental observations (\(\sim 0.5\)--\(0.6\ \mu_{\mathrm{B}}\)) at \(U = 4\ \mathrm{eV}\).

The ground state magnetic structure of \ce{KAgF3} was identified as A-type antiferromagnetism (A-AFM), characterized by uniform magnetic moments on the \ce{Ag^{2+}} ions within the \(ab\)-plane, which are ferromagnetically aligned, while moments are antiferromagnetically coupled along the \(c\)-axis. Concurrently, its orbital order adopts a C-type pattern, with neighboring \ce{Ag^{2+}} orbitals within the \(ab\)-plane aligned orthogonally, whereas along the \(c\)-axis, the orbitals are parallel. Conversely, the ground state of \ce{K2AgF4} was found to be a magnetic phase labeled AFM2, where \ce{Ag^{2+}} ions possess antiparallel spin moments, with their orbital order displaying parallelism amongst neighboring ions. These results can be consistently interpreted within the framework of the Goodenough--Kanamori rules~\cite{GKR}, which describe the relation between orbital ordering and superexchange interactions, thereby providing a microscopic understanding of the observed magnetic and orbital ground states.

\section{method and details}
The calculations were carried out using the BSTATE~\cite{BSTATE} code, within the ultra-soft pseudopotential plane-wave method. For the exchange-correlation potential we used the GGA~\cite{16} given by Perdew-Burke-Ernzerhof(PBE)~\cite{17}. The crystal structures used in the calculations were adopt from  the experiment results~\cite{KAgF3,AgF2009}.  The crystal structures of KAgF$_{3}$ and K$_{2}$AgF$_{4}$ are shown in Fig.1. KAgF$_{3}$ has space group Pnma(\#62) and the lattice constants a=6.3606{\AA}, b=8.2358{\AA}, c=6.0675{\AA} at temperature 2K and a=6.1891{\AA}, b=8.2708{\AA}, c=6.2425{\AA}. K$_{2}$AgF$_{4}$ has space group Cmca(\#64) and lattice constants a=6.1820{\AA}, b=12.63{\AA}, c=6.436{\AA}. We put the long crystal axis-b along z-axis as shown in the Fig.1. Where we defined the x, y, z axes as the [1$\bar{1}$0], [110], and [001] directions of the unit cell respectively. In KAgF$_{3}$, four Ag atoms are defined as Ag$_{1}$(0,0.5,0) and Ag$_{2}$(0.5,0,0) in the first layer, Ag$_{3}$(0,0.5,0.5) and Ag$_{4}$(0.5,0,0.5) in the second layer. The fluorine atoms in the a-b plane are denoted as F$_{pl}$, at the apex of the AgF$_{6}$ octahedra the fluorine atoms are named as F$_{ap}$. In K$_{2}$AgF$_{4}$, two Ag atoms are defined as Ag$_{1}$(0,0,0) and Ag$_{2}$(0.5,0.5,0) in the first layer. 

In order to find the true ground state of KAgF$_{3}$, five different cases were studied. These are non-magnetic(NM), ferromagnetic(FM) and three distinct spin configurations. The first one is A-type antiferromagnetic(A-AFM), where the spins of Ag$^{2+}$ are parallel in a-b plane and antiparallel along c-axis. The second one is C-type antiferromagnetic(C-AFM), where the spins of Ag$^{2+}$ are parallel along c-axis and antiparallel in a-b plane. The third one is G-type antiferromagnetic(G-AFM), where the spins of Ag$^{2+}$ are antiparallel to all the nearest neighbors. To investigate the true ground state of K$_{2}$AgF$_{4}$, four different cases were considered. These are nonmagnetic(NM) and three distinct spin states of FM, interlayer-antiferromagnetism(AFM1), intralayer-antiferromagnetism(AFM2).
The plane wave energy cutoff was set to 340eV for all states. The brillouin zone integrations were performed with a $15\times15\times15$ special k-points grid.  The electron correlations of Ag-4d states were treated by the GGA+U~\cite{HU,ldau} method. Due to the  localized nature of Ag-4d electronic states in the material, we employed the GGA+U method~\cite{HU,ldau} with U varying from 0 to  4 eV.

\begin{figure}[htbp]
\begin{center} 
\includegraphics[clip,scale=0.45]{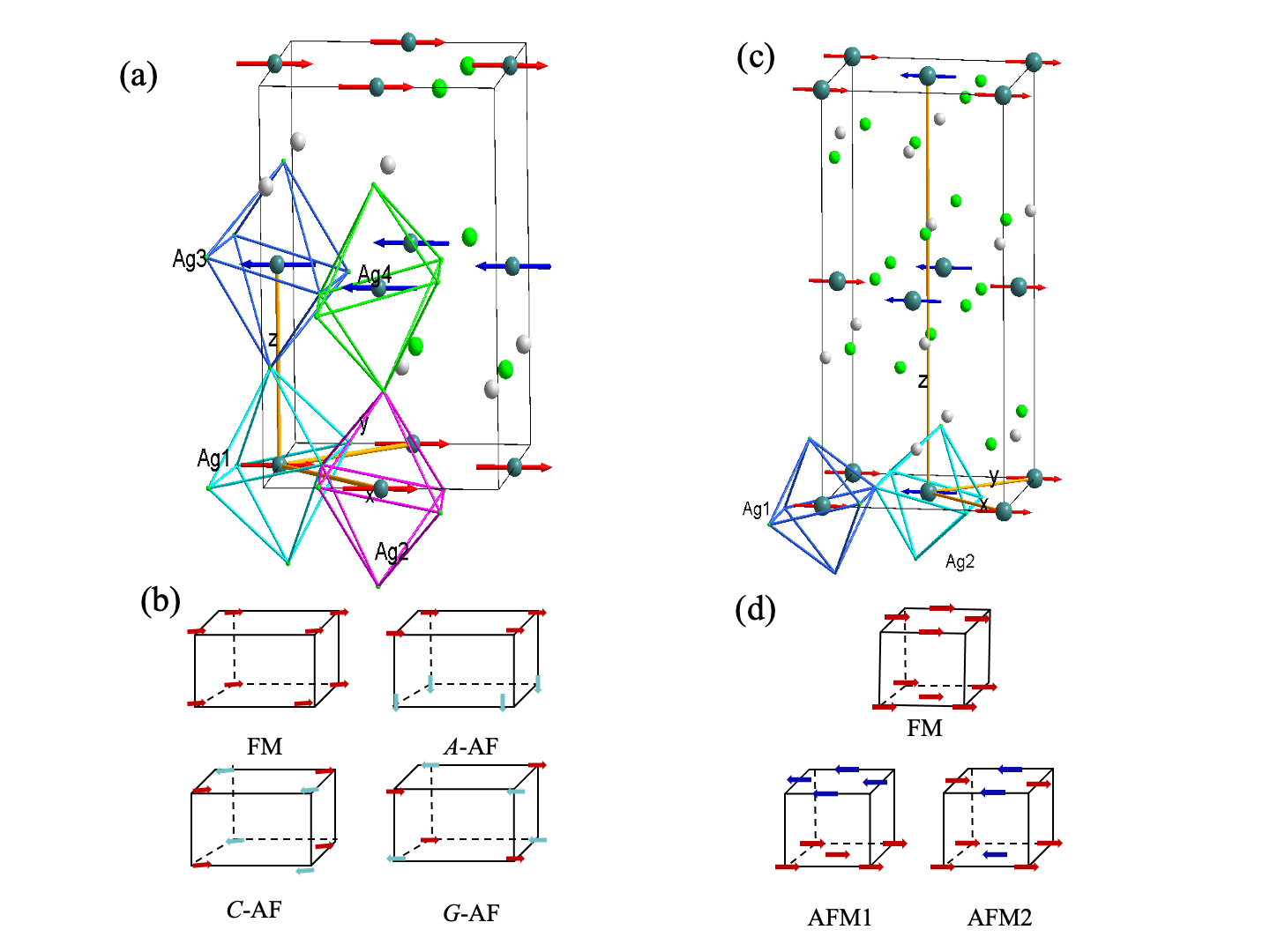}
 \caption{ (a) The crystal structure of KAgF$_3$, we defined the x, y, z axes as the [1$\bar{1}$0], [110], and [001] directions of the unit cell respectively. (b) The FM, A-AF, C-AF, and G-AF magnetic configurations. (c) The crystal structure of K$_2$AgF$_4$. (d) The FM, AFM1, AFM2 magnetic configurations.  }
\label{structure}
\end{center}
\end{figure}

The direct observations of anisotropic Fermi surface and spin-splitting band structure are difficult. However the measurements of anisotropic optical conductivity (AOC), anomalous Hall, Nernst, and thermal Hall effects, along with magneto-optical responses including the Kerr and Faraday effects have been proved as a useful way to investigate the characters of alter-magnetism  indirectly~\cite{Chen_2024-Mn-rev,layer-nernst}. These properties   are calculated from the converged Kohn-Sham wave functions $|\psi_{n\bf k}\rangle$ and eigenvalues
$E_n({\bf k})$ by using the following Kubo formula~\cite{opteq1,opteq2,KCrF}:
\begin{eqnarray}
\sigma_{\alpha\beta}(\omega)
&=&-\frac{16}{V}\sum_{\bf k\it n}if_{n\bf k}\sum_{m}
\frac{1}{\omega_{mn}^2-(\omega+i\delta)^2} \nonumber\\
&&\left[\frac{\omega+i\delta}{\omega_{mn}}
Re(\pi_{nm}^\alpha\pi_{mn}^\beta)+iIm(\pi_{nm}^\alpha\pi_{mn}^\beta)
\right]
\end{eqnarray}
where $\alpha$ and $\beta$ (=$x,y,z$) are indices for directions,
$\omega$ is the excitation energy, $V$ is the volume of the unit cell,
$n$ and $m$ are band indices, $f_{n\bf k}$ is the Fermi distribution
function, $\omega_{mn}=E_{m}({\rm k})-E_{n}({\rm k})$ and $\delta$ is
the lifetime broadening ($\delta$=0.01Ry in this work),
$\pi_{nm}^\alpha=\langle \psi_{n\bf k}|(-i\nabla_\alpha)|\psi_{m\rm
k}\rangle$ are the matrix elements of the momentum operator.

\begin{eqnarray}
	\Omega_n^\gamma(\mathbf{k}) &=& -2\,\mathrm{Im}\sum_{m\neq n}
	\frac{\pi_{nm}^\alpha \pi_{mn}^\beta}{(\varepsilon_m - \varepsilon_n)^2}, \label{Berry}\\
	\sigma_{\alpha\beta}^{\mathrm{AHC}} (\mu)&=& -\frac{e^2}{\hbar}\int \frac{d^3\mathbf{k}}{(2\pi)^3}f(\mu) \Omega^{\gamma}(\mathbf{k}), \label{AHC}\\
	\alpha_{\alpha\beta}^{\mathrm{ANC}}(\mu) &=& \int \frac{\varepsilon - \mu}{eT}
	\left(-\frac{\partial f}{\partial \varepsilon}\right)
	\sigma_{\alpha\beta}^{\mathrm{AHC}}\,d\varepsilon, \label{ANC}\\
	\kappa_{\alpha\beta}^{\mathrm{TAHC}}(\mu) &=& \int \frac{(\varepsilon - \mu)^2}{e^2T}
	\left(-\frac{\partial f}{\partial \varepsilon}\right)
	\sigma_{\alpha\beta}^{\mathrm{AHC}}\,d\varepsilon, \label{TAHC}
\end{eqnarray}

Where $\Omega^{\gamma}(\mathbf{k})$ is the momentum-resolved Berry curvature; $\alpha$, $\beta$, and $\gamma$ denote the Cartesian indices corresponding to the $x$, $y$, and $z$ directions; $\mu$ is the chemical potential, and $f = 1/\left[\exp\left((\varepsilon - \mu)/k_BT\right) + 1\right]$ is the Fermi--Dirac distribution.

For the complex Kerr and Faraday angle, we adopt a simplified expression under the assumption of a small rotation angle~\cite{mokeeq1,mokeeq2,mokeeq3,mokeeq4,mokeeq5,mokeeq6,mokeeq7,mokeeq8,mokeeq9}: 
\begin{eqnarray}
    \phi^{z}_{K} = \vartheta^{z}_{K} + i\varepsilon^{z}_{K} \approx \frac{-\nu_{xyz} \sigma_{xy}}{\sigma_{0} \sqrt{1 + i(4\pi/\omega)\sigma_{0}}} ,  \\
    \phi^{z}_{F} = \vartheta^{z}_{F} + i\varepsilon^{z}_{K} \approx \frac{-\nu_{xyz} \sigma_{xy}}{ \sqrt{1 + i(4\pi/\omega)\sigma_{0}}} \frac{2\pi}{c} , 
\end{eqnarray}
where $\sigma_0 = (\sigma_{xx} +\sigma_{yy})/2$.

\section{Results and Discussion}

\subsection{Results of KAgF$_3$}

We first investigated the electronic structure of \ce{KAgF3} in the nonmagnetic (NM) state. As shown in Fig.2, the crystal field of the distorted \ce{AgF6} octahedron splits the \ce{Ag^{2+}} $4d^9$ orbitals into  fully occupied $t_{2g}$ ($d_{yz}$, $d_{xy}$, $d_{xz}$) and $d_{3x^2-r^2}$) states, and  partially occupied $d_{z^2-y^2}$ orbital.  Furthermore, a strong hybridization is observed between the F $2p$ and Ag $4d$ orbitals.
Given the experimentally established antiferromagnetic insulating ground state of \ce{KAgF3}, whose underlying physical mechanism remains incompletely understood, we have systematically evaluated its various magnetic orderings. These include the ferromagnetic (FM), along with the A-, C-, and G-type antiferromagnetic (A-AFM, C-AFM, G-AFM) configurations, whose spin structures are schematically depicted in Fig. 1. To account for the strong electron correlations in the Ag $4d$ states of \ce{KAgF3}, we employed the GGA+$U$ method. The total energies and local magnetic moments for various magnetic orders were calculated with the Hubbard $U$ parameter ranging from 0 to 4 eV. 

At $U = 0$, all magnetic configurations converge to a non-magnetic (NM) state. The local spin moment on Ag increases monotonically with the Hubbard $U$ parameter, and the A-type antiferromagnetic (A-AFM) state emerges as the ground state. At $U = 4$~eV, the calculated Ag magnetic moment reaches $\sim 0.5~\mu_{\mathrm{B}}$, in good agreement with experimental values.

To quantitatively determine the exchange interactions within the $ab$-plane ($J_{ab}$) and along the $c$-axis ($J_{c}$), we map the total energies of different magnetic states---ferromagnetic (FM, $E_{\mathrm{F}}$), A-AFM ($E_{\mathrm{A}}$), C-AFM ($E_{\mathrm{C}}$), and G-AFM ($E_{\mathrm{G}}$)---onto a classical Heisenberg model. The nearest-neighbor exchange coupling constants are derived as follows:
\begin{equation}
\begin{aligned}
J_c &= \frac{E_{\mathrm{F}} - E_{\mathrm{G}} - E_{\mathrm{A}} + E_{\mathrm{C}}}{4S^2} \\
J_{ab} &= \frac{E_{\mathrm{F}} - E_{\mathrm{G}} + E_{\mathrm{A}} - E_{\mathrm{C}}}{8S^2}
\label{Heisen}
\end{aligned}
\end{equation}
where  S is the moment of \ce{Ag^{2+}}. 

\begin{figure}
\includegraphics[clip,scale=0.45]{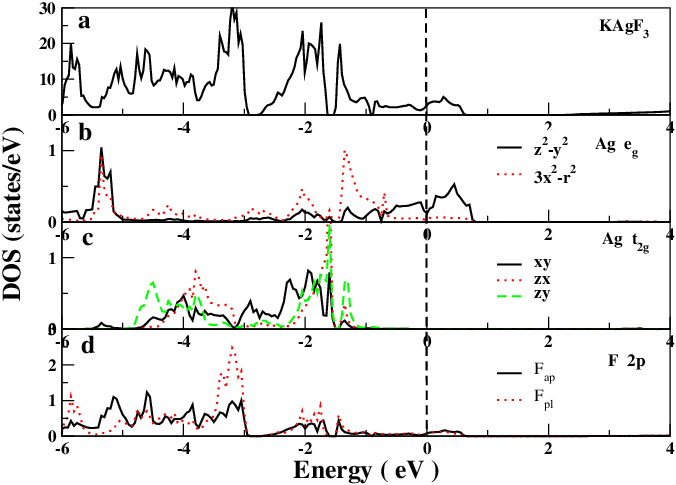}
\label{113dos}
\caption{(a) The total density states of KAgF$_3$ and the projected density states of Ag-e$_{g}$ orbital (b) , Ag-t$_{2g}$ orbital (c)  and F-2p (d).}
\end{figure}

\begin{figure}
    \centering
    \includegraphics[clip,scale=0.3]{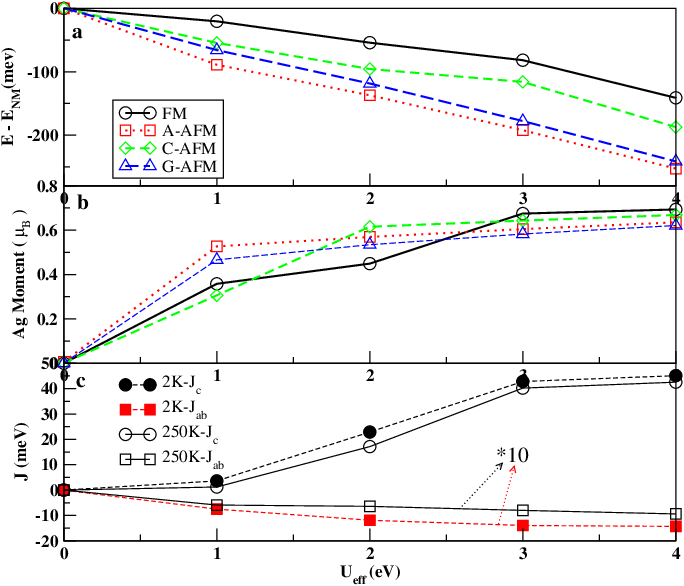}
    \caption{(a) The total energies (relative to the NM state) and (b) magnetic moments of different spin configurations in KAgF$_{3}$. (c) The calculated exchange interactions for \ce{KAgF3} at temperatures of 2 K and 250 K.}
    \label{EMJ}
\end{figure}

\begin{figure}
    \centering
    \includegraphics[clip,scale=0.35]{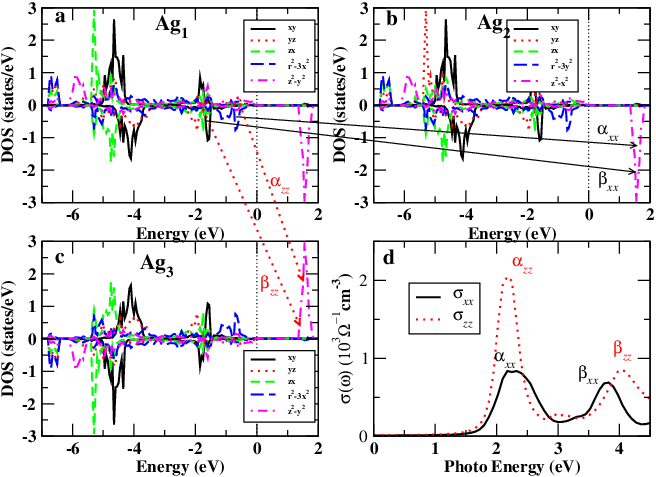}
    \caption{The PDOS of Ag$_{1}$(a), Ag$_{2}$(b), Ag$_{3}$(c) and the optical conductivity (d) of KAgF$_{3}$ calculated with GGA+U (U$_{\text{eff}}$ = 4.0 eV) method. $a_{xx}$ denotes the optical conductivity within the $ab$-plane, while $a_{zz}$ represents the optical conductivity along the $c$-axis. The line and arrow indicate the charge transition.}
    \label{113-A-pdos}
\end{figure}

\begin{figure}
    \centering
    \includegraphics[clip,scale=0.4]{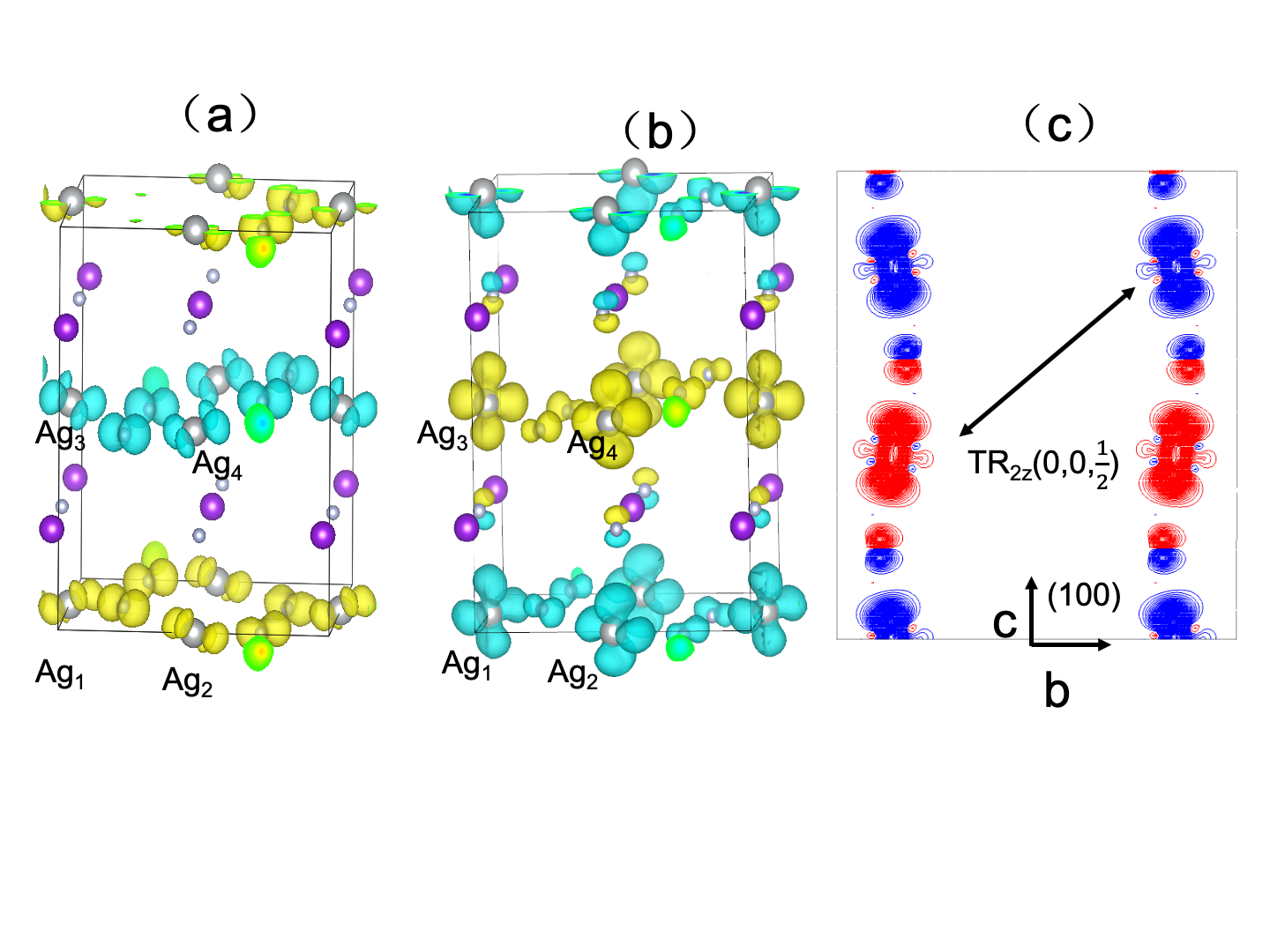}
    \caption{(a) The charge density of both spin-up and spin-down states in the range of -1.4 eV to 0.0 eV (E$_{f}$) for A-AFM. (b) The charge density of the unoccupied state in the range of 0.0 eV to 2.0 eV, with spin-up in yellow and spin-down in green. (c) The contours of the charge density on the (100) plane.}
    \label{113-A-charge}
\end{figure}

From Eq.\eqref{Heisen}, we obtain the exchange interactions $J_{ab}$ and $J_{c}$, as plotted in Fig.~3(c). It can be seen that $J_{ab}$ is negative while $J_{c}$ is positive, indicating that the magnetic moments of \ce{Ag^{2+}} favor parallel alignment within the $ab$-plane and antiparallel alignment along the $c$-axis, consistent with the A-AFM ground state. At the temperature both 2 K and 250 K, 
the abslute value of $J_{c}$  is about 10 times larger than that of $J_{ab}$. Such significant difference can be understood with the help of its partial density of state in the Fig.\ref {113-A-pdos}  and the charge density of Fig.\ref{113-A-charge}. 

\begin{table}[h]
    \centering
    \caption{
        Symmetry operations for the orthorhombic \textit{Pnma} space group (No.~62) and their classification under different antiferromagnetic (AFM) configurations. The crystallographic symmetry group $G$ comprises eight operations. For each AFM configuration, these are categorized into: (1) four operations ($G-A$) that preserve the spin channel, acting within a single spin subsystem; and (2) four spin-flip operations ($\hat{T}A$) mediated by the time-reversal operator $\hat{T}$.
    }
    \label{tab:SY}
    \begin{tabular}{|c|c|c|}
        \hline
        \multicolumn{3}{|c|}{Group $G$: $E$, $I$, $\{R_{2x}|\frac{1}{2}\frac{1}{2}0\}$, $\{R_{2y}|\frac{1}{2}\frac{1}{2}\frac{1}{2}\}$, $\{R_{2z}|00\frac{1}{2}\}$, $\{M_{x}|\frac{1}{2}\frac{1}{2}0\}$, $\{M_{y}|\frac{1}{2}\frac{1}{2}\frac{1}{2}\}$, $\{M_{z}|00\frac{1}{2}\}$} \\
        \hline
        AFM Type & $G-A$ (Spin-preserving) & $\hat{T}A$ (Spin-flip) \\
        \hline
        $A$-AFM & 
        $E$, $I$, $\{R_{2x}|\frac{1}{2}\frac{1}{2}0\}$, $\{M_{x}|\frac{1}{2}\frac{1}{2}0\}$ &
        $\hat{T}\{R_{2y}|\frac{1}{2}\frac{1}{2}\frac{1}{2}\}$, $\hat{T}\{M_{y}|\frac{1}{2}\frac{1}{2}\frac{1}{2}\}$, $\hat{T}\{R_{2z}|00\frac{1}{2}\}$, $\hat{T}\{M_{z}|00\frac{1}{2}\}$ \\
        \hline
        $C$-AFM & 
        $E$, $I$, $\{R_{2z}|00\frac{1}{2}\}$, $\{M_{z}|00\frac{1}{2}\}$ &
        $\hat{T}\{R_{2x}|\frac{1}{2}\frac{1}{2}0\}$, $\hat{T}\{R_{2y}|\frac{1}{2}\frac{1}{2}\frac{1}{2}\}$, $\hat{T}\{M_{x}|\frac{1}{2}\frac{1}{2}0\}$, $\hat{T}\{M_{y}|\frac{1}{2}\frac{1}{2}\frac{1}{2}\}$ \\
        \hline
        $G$-AFM & 
        $E$, $I$, $\{R_{2y}|\frac{1}{2}\frac{1}{2}\frac{1}{2}\}$, $\{M_{y}|\frac{1}{2}\frac{1}{2}\frac{1}{2}\}$ &
        $\hat{T}\{R_{2x}|\frac{1}{2}\frac{1}{2}0\}$, $\hat{T}\{M_{x}|\frac{1}{2}\frac{1}{2}0\}$, $\hat{T}\{R_{2z}|00\frac{1}{2}\}$, $\hat{T}\{M_{z}|00\frac{1}{2}\}$ \\
        \hline
    \end{tabular}
\end{table}

Fig.\ref {113-A-pdos}  shows the partial density of states (PDOS) of Ag$_{1}$, Ag$_{2}$, and Ag$_{3}$ in KAgF$_{3}$, along with the calculated optical conductivity. In KAgF$_{3}$, the Ag$_{1}$-F$_{pl}$ distance is 2.4152~\AA\ along the $x$-axis and 2.0991~\AA\ along the $y$-axis, while the Ag$_{1}$-F$_{ap}$ distance along the $z$-axis is 2.117~\AA. Due to the above Jahn-Teller (JT) distortion and the rotation of the [AgF$_{6}$]$^{4-}$ octahedra, the degenerate energy level $t_{2g}^{\uparrow}$ of Ag$_{1}$ splits according to $E_{zy} > E_{xy} > E_{zx}$, where $E_{zy}$, $E_{xy}$, and $E_{zx}$ represent the energies of the $d_{zy}$, $d_{xy}$, and $d_{zx}$ orbitals, respectively. Also because of the JT, the $e_{g}^{\downarrow}$ level splits as $E_{z^{2}-y^{2}} > E_{r^{2}-3x^{2}}$, where $E_{z^{2}-y^{2}}$ and $E_{r^{2}-3x^{2}}$ are the energy levels of the $d_{z^{2}-y^{2}}$ and $d_{r^{2}-3x^{2}}$ orbitals. 
There are nine electrons on the \ce{Ag^{2+}} ion. Six of these electrons first occupy the spin-up and spin-down $t_{2g}^{\uparrow\downarrow}$ orbitals. This is because the crystal field of the [AgF$_{6}$] octahedron lowers the energy of the $t_{2g}$ orbitals and raises the energy of the $e_g$ orbitals. The remaining three electrons then occupy the higher-energy four $e_g$ orbitals. Specifically, for the Ag$_{1}$ ion, electrons first fill the $d_{r^{2}-3x^{2}}^{\uparrow\downarrow}$ orbital, followed by the $d_{z^{2}-y^{2}}^{\uparrow}$ orbital, leaving the $d_{z^{2}-y^{2}}^{\downarrow}$ orbital empty, as shown in the Fig.\ref {113-A-pdos}a and Fig.\ref{113-A-charge}b.

For Ag$_{2}$, the corresponding bond lengths are Ag$_{2}$-F$_{pl}$($x$) = 2.0991~\AA, Ag$_{2}$-F$_{pl}$($y$) = 2.4152~\AA, and Ag$_{2}$-F$_{ap}$($z$) = 2.117~\AA. As a result of the Jahn–Teller distortion, the $e_{g}^{\downarrow}$ level splits into $E_{z^{2}-x^{2}} > E_{r^{2}-3y^{2}}$, where $E_{z^{2}-x^{2}}$ and $E_{r^{2}-3y^{2}}$ correspond to the energy levels of the $d_{z^{2}-x^{2}}$ and $d_{r^{2}-3y^{2}}$ orbitals, respectively. Consequently, the remaining down-spin electron occupies the lower-energy $d_{r^{2}-3y^{2}}^{\downarrow}$ state, leaving the $d_{z^{2}-x^{2}}^{\downarrow}$ orbital empty.

Therefore, Ag$_{1}$ and Ag$_{3}$ share the same orbital ordering, whereas Ag$_{1}$ and Ag$_{2}$ exhibit a zig-zag pattern, as illustrated in Fig.~\ref{113-A-pdos} and Fig.~\ref{113-A-charge}. The orbital overlap between Ag$_{1}$ and Ag$_{3}$ is significantly larger than that between Ag$_{1}$ and Ag$_{2}$, leading to a larger exchange coupling $J_c$ compared to $J_{ab}$ (Fig.\ref{EMJ}). Meanwhile, according to the Goodenough–Kanamori rules~\cite{GKR}, the exchange coupling between parallel orbitals is antiferromagnetic, whereas the coupling between perpendicular orbitals is  ferromagnetic (Fig.\ref{113-A-pdos} and Fig.\ref{113-A-charge}).

In conventional antiferromagnets, the symmetry operations relating two antiparallel magnetic ions are either space inversion combined with time reversal or translation combined with time reversal. In the A-AFM phase of \ce{KAgF3}, however, the relevant symmetry operation is a combination of rotation and time reversal, as summarized in Table~\ref{tab:SY} and schematically depicted in Fig.~\ref{113-A-charge}(c). 

\begin{figure}
\begin{center}
\includegraphics[clip,scale=0.4]{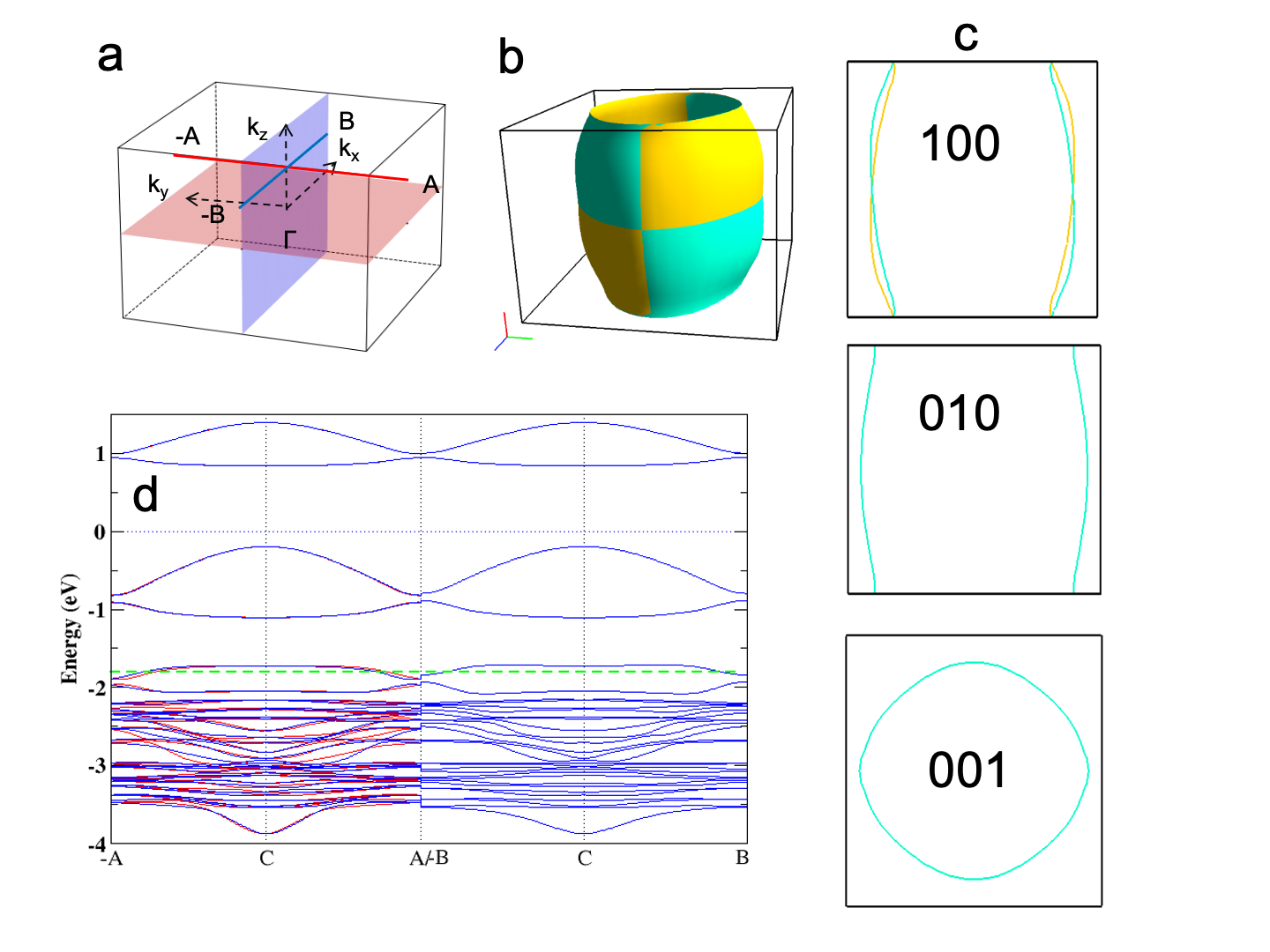}
\end{center} 
\caption{(a) The Brillouin zone, hight symmetry planes and high symmetry lines. (b) The 3D Fermi surface at the energy -1.75 eV.  (c)  The cut section of the Fermi surface at (100), (010) and (001) planes. (d) The band structure along the high symmetry lines }.
\label{BZ}
\end{figure}

Owing to the symmetry operations $\hat{T}\{M_{y}|\frac{1}{2}\frac{1}{2}\frac{1}{2}\}$ and $\hat{T}\{M_{z}|00\frac{1}{2}\}$, the spin-up and spin-down bands remain degenerate on  the $K_y=0$ and $K_z=0$ planes, respectively, as indicated by the blue and orange planes in Fig.~\ref{BZ}(a). The Fermi surface at an energy of $-1.75$ eV is shown in Fig.~\ref{BZ}(b), with a corresponding cross-section displayed in Fig.~\ref{BZ}(c). At the $K_x=0$ plane, spin-up and spin-down FS sheets are splitting, because of lacking symmetry operation to protect them. While the Fermi surface sheets remain degenerate at the $K_y=0$ and $K_z=0$ planes, protected by the $\hat{T}\{M_{y}|\frac{1}{2}\frac{1}{2}\frac{1}{2}\}$ and $\hat{T}\{M_{z}|00\frac{1}{2}\}$ symmetries, respectively. Furthermore, the band structure in Fig.\ref{BZ}(d) exhibits spin splitting along the A-C-A high-symmetry line, while spin degeneracy is preserved along the B-C-B path.

For compounds with $\hat{\mathcal{PT}}$ symmetry, the Berry curvature vanishes in all directions due to the constraint $\hat{\mathcal{PT}}\Omega_n^{\alpha}(\mathbf{k}) = -\Omega_n^{\alpha}(\mathbf{k}) \implies \Omega_n^{\alpha}(\mathbf{k}) \equiv 0$~\cite{Topandoptical}. In contrast, the altermagnetic A-AFM KAgF${3}$ breaks $\hat{\mathcal{PT}}$ symmetry while retaining the eight symmetry operations listed in Table~\ref{tab:SY}, such as \{${R{2x}|\frac{1}{2}\frac{1}{2}0}$\}. The operation $R_{2x}$ imposes $R_{2x}\Omega_n^{x}(\mathbf{k}) = \Omega_n^{x}(k_x,-k_y,-k_z) = +\Omega_n^{x}(\mathbf{k})$, leading to a nonvanishing $\Omega_n^{x}(\mathbf{k})$. Meanwhile, the constraints $\hat{\mathcal{T}}R_{2y}\Omega_n^{y}(\mathbf{k}) = -\Omega_n^{y}(\mathbf{k}) = 0$ and $\hat{\mathcal{T}}R_{2z}\Omega_n^{z}(\mathbf{k}) = -\Omega_n^{z}(\mathbf{k}) = 0$ force $\Omega_n^{y/z}(\mathbf{k})$ to be trivial.

\begin{figure}
\begin{center}
\includegraphics[clip,scale=0.4]{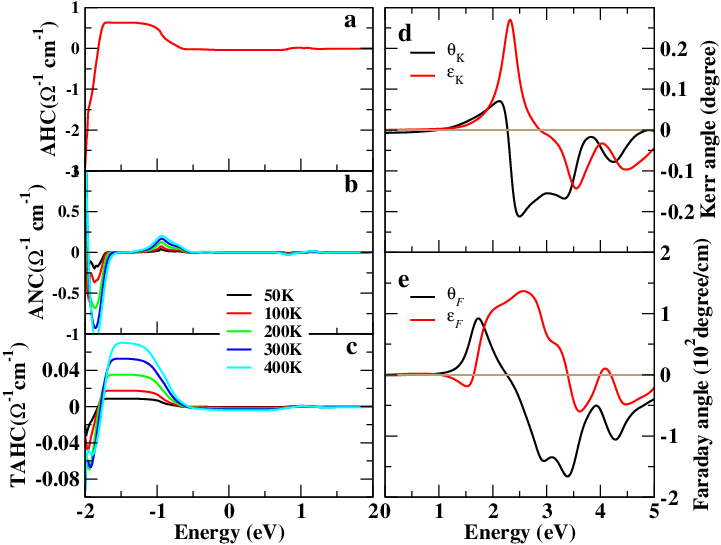}
\end{center} \caption{(a) The  anomalous Hall conductivity (AHC). (b) anomalous Nernst conductivity (ANC) and (c) anomalous thermal Hall conductivity (TAHC) varying with  temperature. (d) The magneto-optical Kerr rotation $\theta_k$  and Kerr ellipticity $\epsilon_k$. (e) Faraday rotations  angle $\theta_F$ and Faraday ellipticity $\epsilon_F$. }.
\end{figure}

In the case of $A$-AFM KAgF$_3$, only the $\Omega_n^{x}$-component is nonzero, inducing anomalous Hall conductivity (AHC), anomalous Nernst conductivity (ANC), and thermal anomalous Hall conductivity (TAHC), as shown in Fig.~7. Between –2 and –0.5~eV, the AHC, ANC, and TAHC (Fig.~7a–c) vary significantly with Fermi energy shift, whereas they remain nearly unchanged and negligibly small in the energy range from 0 to 2~eV. This indicates that the anomalous Hall and Nernst effects are readily observable in hole-doped compounds, but are strongly suppressed in electron-doped samples. As shown in Fig.7d, for incident light with a photon energy around 2.5 eV, the Kerr rotation angle upon reflection from the surface of $A$-AFM KAgF$_3$ can reach up to 0.2$^\circ$ ($\theta_K$). Simultaneously, the material converts the linearly polarized light into elliptically polarized light with an ellipticity of approximately 0.2. When the  linearly polarized light transfer though the KAgF$_3$, the 

In conventional antiferromagnets the operations with preserved $\hat{\mathcal{P}\mathcal{T}}$ symmetry, the Berry curvature vanishes identically in all momentum directions due to the constraint $\hat{\mathcal{P}\mathcal{T}}\Omega_n^{\alpha}(\mathbf{k}) = -\Omega_n^{\alpha}(\mathbf{k}) \implies \Omega_n^{\alpha}(\mathbf{k}) \equiv 0$~\cite{Topandoptical}. In contrast, the altermagnetic $G$-type antiferromagnet NaCoF$_3$ breaks $\hat{\mathcal{P}\mathcal{T}}$ symmetry while retaining the eight symmetry operations listed in Table. II, with the residual symmetries enforcing anisotropic Berry curvature components through $\mathcal{R}_{2y}\Omega_n^{y}(\mathbf{k}) = \Omega_n^{y}(-k_x, k_y, -k_z) = +\Omega_n^{y}(\mathbf{k})$ and $\hat{\mathcal{T}}\mathcal{R}_{2x/z}\Omega_n^{x/z}(\mathbf{k}) = -\Omega_n^{x/z}(\mathbf{k}) = 0$, yielding nonvanishing $\Omega_n^{y}(\mathbf{k})$ but vanishing $\Omega_n^{x/z}(\mathbf{k})$. The altermagnetic NaCoF$_3$ exhibits exclusively $y$-direction anomalous transport properties, manifested in three distinct phenomena: the anomalous Hall conductivity (AHC), anomalous Nernst conductivity (ANC), and thermal anomalous Hall conductivity (TAHC). All of them display pronounced peaks centered around 4 eV in their respective spectra as clearly shown in Fig.7(a)-(c).

The magneto-optical response exhibits distinctive signatures characteristic of alter-magnetism~\cite{yaoMOKE,Topandoptical}, with the Kerr rotation angle $\theta_K$ and ellipticity $\epsilon_K$ attaining maximum values of $9\times10^{-2}$degree and  $8\times10^{-2}$degree(Fig.7d) respectively. While the Kerr rotation angle is smaller than that of ferromagnetic compounds~\cite{Fe2MnSn,CoTiSn,RuF4-Kerr}, approximately 10$\sim$50 times larger than that of $\hat{\mathcal{P}\mathcal{T}}$-symmetry-protected antiferromagnetic materials~\cite{mokeeq1,mokeeq2,mokeeq3,MnBiTe}, and comparable to other altermagnetic compounds~\cite{yaoMOKE,alterKerr1,NANO-Kerr}. These pronounced Kerr effects serve as a quantitative probe of the altermagnetic state through polarization-resolved reflectivity measurements. Furthermore, transmission magneto-optical response reveal an exceptionally large Faraday rotation\cite{Topandoptical} of $1.1\times10^3$ degree/cm (Fig.7e), which exceeds typical values observed in conventional magnetic materials by nearly an order of magnitude~\cite{yaoMOKE,alterKerr1,NANO-Kerr}. The coexistence of these robust magneto-optical responses demonstrates significant potential for applications in altermagnetism-based optical spintronic technologies.

\subsection{Results of K$_2$AgF$_4$}

K$_2$AgF$_4$ is another AgF-based compound that exhibits strong Jahn–Teller distortion and shares the $4d^9$ electronic configuration of Ag$^{2+}$ with KAgF$_3$, yet it crystallizes in a distinct quasi-two-dimensional structure. Unlike the \textit{Pnma} phase of KAgF$_3$, where AgF$_6$ octahedra are connected in a 3D corner-sharing network, K$_2$AgF$_4$ forms layers in the \textit{ab}-plane via in-plane corner-sharing between adjacent AgF$_6$ octahedra. Our calculations considered three magnetic configurations---FM, AFM1, and AFM2---as illustrated in Fig.~1(b), and identified the in-plane antiferromagnetic order (AFM2) as the magnetic ground state at $U_{\text{eff}} = 4$~eV. We therefore focus on the AFM2 phase in the following analysis. In this magnetic structure, all crystal symmetry operations connect atoms with parallel spins, and the combined $R\hat{\mathcal{T}}$ symmetry (spin-flip operation) is absent. As a result, the anomalous Hall conductivity (AHC) vanishes.

\begin{figure}[htbp]
    \centering
    \includegraphics[clip,scale=0.4]{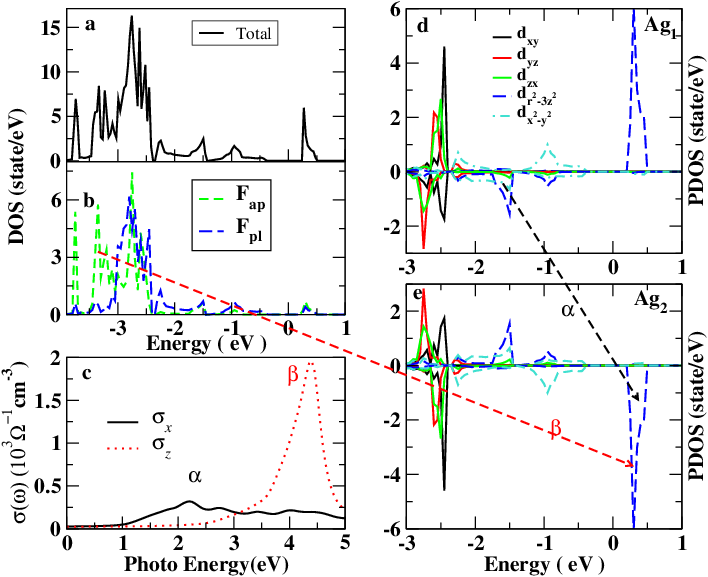}
    \caption{Properties of K$_2$AgF$_4$: 
        \textbf{(a)} Total density of states (DOS).
        \textbf{(b)} Projected DOS (PDOS) of F atoms.
        \textbf{(c)} Optical conductivity ($\sigma_{x}$: in-plane; $\sigma_{z}$: out-of-plane). Arrows mark the key electronic transitions.
        \textbf{(d)} PDOS of the Ag$_1$ atom.
        \textbf{(e)} PDOS of the Ag$_2$ atom.
    }
    \label{214-pdos}
\end{figure}

\begin{figure}[htbp]
    \centering
    \includegraphics[clip,scale=0.4]{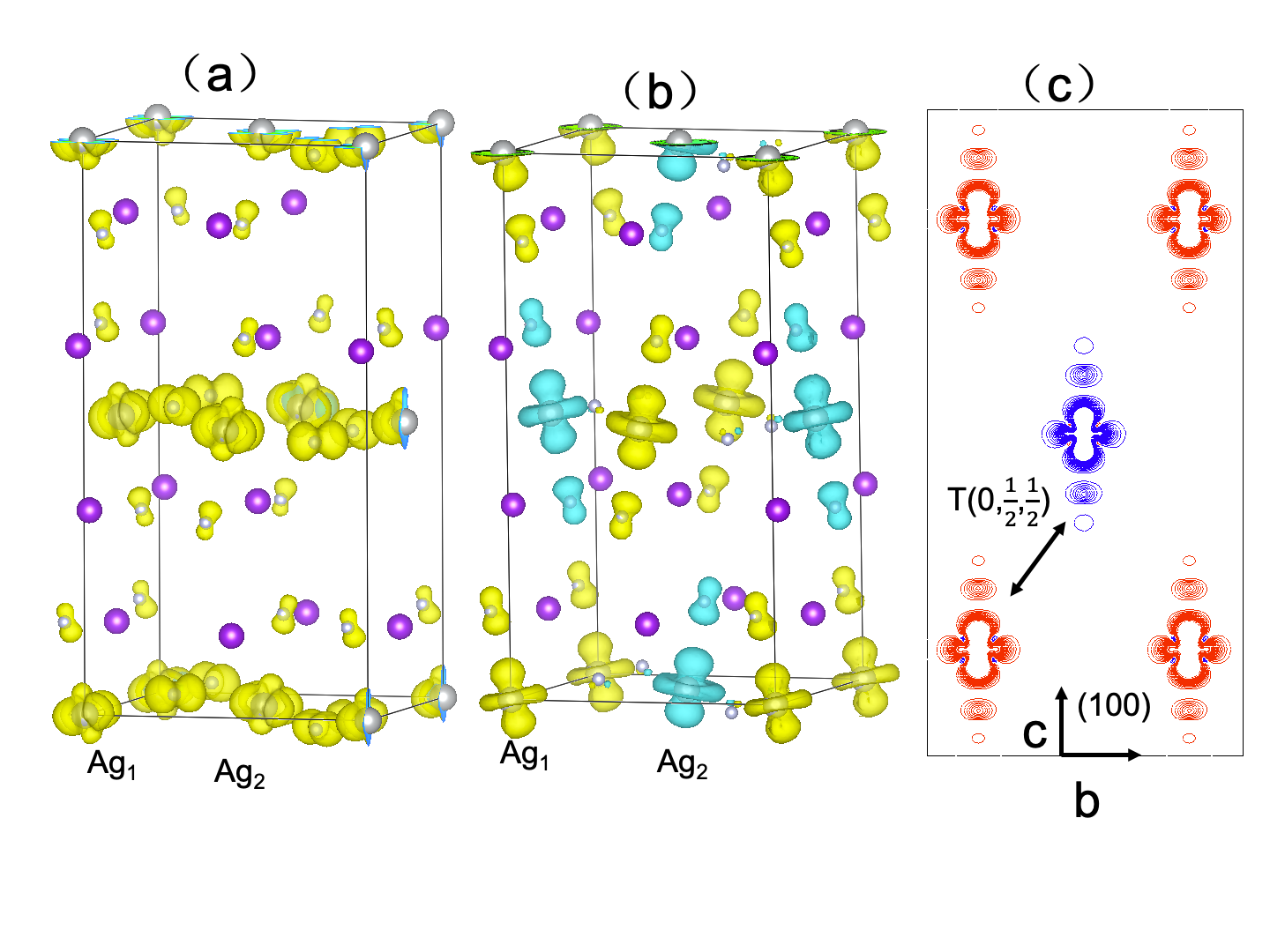}
    \caption{Charge density analysis of the A-AFM state. (a) Charge density of both spin-up and spin-down states within the energy window from $-1.4$ eV to the Fermi level ($E_F$). (b) Charge density of the unoccupied states from $0.0$ eV to $2.0$ eV, with the spin-up and spin-down components depicted in yellow and green, respectively. (c) Contour plot of the charge density on the (100) plane for the charge in (b).}
    \label{214-charge}
\end{figure}

The electronic structure and optical properties are summarized in Fig.\ref{214-pdos}. The atom-projected density of states reveals that the electronic states within $\pm 1$~eV of the Fermi level are primarily derived from Ag atoms (panels d-e). At lower energies, between $-4$ and $-2$~eV, the states arise from the hybridization of Ag $d$ orbitals with F $p$ orbitals (panels b, d-e). It is noteworthy that the unoccupied orbitals of both Ag1 and Ag2 share the $d_{r^{2}-3z^{2}}$ character but differ in their spin configurations. A marked anisotropy is observed in the optical conductivity (panel c). The in-plane ($a$-$b$ plane) component (black solid line $\sigma_x$) shows a pronounced peak at approximately 2~eV, whereas the out-of-plane component  (red dashed line  $\sigma_z$) features a distinct peak near 4.5~eV, underscoring the quasi-two-dimensional nature of the material's optical response.

The in-plane optical conductivity peak $\alpha$ originates from a $d$-$d$ transition between the $Ag^{\uparrow}_{1}$-$d_{r^{2}-3z^{2}}$ and $Ag^{\downarrow}_{2}$-$d_{r^{2}-3z^{2}}$ orbitals, as indicated by the black dashed arrow connecting panels (d) and (e). In contrast, the out-of-plane peak $\beta$ arises from a $p$-$d$ charge-transfer transition from the F$_{\text{ap}}$-$p$ orbital to the $d_{r^{2}-3z^{2}}$ orbital, marked by the red dashed arrow between panels (b) and (d). The significantly greater intensity of peak $\beta$ is attributed to the strong $p$-$d$ orbital hybridization.

The two-dimensional character of the system, initially indicated in Fig.~\ref{214-pdos}, is further corroborated by the charge density distributions in Fig.~\ref{214-charge}. Panel (a) shows the charge density for spin+down states within an energy window from $-1.4$~eV to the Fermi level ($E_F$), while panel (b) presents the density of unoccupied states (0.0-2.0~eV), with spin-up and spin-down components colored in yellow and green, respectively. The corresponding contour plot on the (100) plane is displayed in panel (c).

As shown in panel (c), the charge density distributions around the two symmetry-related Ag sites (Ag1 and Ag2)—which exhibit opposite spin orientations—are connected by a composite symmetry operation: the product of time-reversal symmetry $\mathcal{T}$ and a lattice translation $\mathbf{t} = (0, 1/2, 1/2)$. This symmetry enforces a vanishing Berry curvature throughout the entire Brillouin zone. Consequently, the anomalous Hall conductivity (AHC) is identically zero in this system.

\subsection{Summary and Conclusion}
Based on first-principles calculations and systematic symmetry analysis, we investigate the electronic, magnetic, and orbital structures of AgF-based compounds KAgF$_3$ and K$_2$AgF$_4$. We find that the magnetic ground state of \textit{Pnma} KAgF$_3$ exhibits A-type antiferromagnetic (A-AFM) order, accompanied by C-type orbital ordering, which can be well explained by the Goodenough–Kanamori–Rules.  Since the A-AFM phase breaks $\mathcal{PT}$ symmetry, it exhibits nonzero Berry curvature and gives rise to  anomalous Hall conductivity, anomalous Nernst conductivity, and thermal anomalous Hall conductivity. At the same time, it presents strong magneto-optical responses, manifesting through pronounced Kerr and Faraday effects. On the other hand, K$_2$AgF$_4$ behaves as a conventional collinear antiferromagnet preserving $\mathcal{PT}$ symmetry, hence precluding the emergence of an anomalous Hall response.

\begin{acknowledgments}
The authors  acknowledge the supports from NSF of China (No.11904084 and No.10947001) and the Innovation Scientists and Technicians Troop Constriction Projects of Henan Province (Grant No. 104200510014).
\end{acknowledgments}

\bibliography{reference111.bib}

\end{document}